\documentclass[aip,rsi,reprint]{revtex4-1} 

\usepackage{siunitx}
\DeclareSIUnit{\sqrthz}{\ensuremath{\sqrt{\text{\hertz}}}}
\DeclareSIUnit{\dbm}{\decibel\text{m}}
\DeclareSIUnit{\dbc}{\decibel\text{c}}

\usepackage{amsmath}
\usepackage{amsfonts}
\usepackage{txfonts}
\usepackage{graphicx}
\usepackage[caption=false]{subfig}
\usepackage{dcolumn}
\usepackage{multirow}
\usepackage{hyperref}

\DeclareGraphicsExtensions{.png,.tiff,.ai,.eps}

\newcommand\T{\rule{0pt}{2.6ex}}       

\graphicspath{{./}{./Figures/}}

\usepackage{natbib}

\begin{document}

\title{A flexible, open-source radio-frequency driver for acousto-optic and electro-optic devices}

\author{D.S. Barker}
\email[]{daniel.barker@nist.gov}
\author{N.C. Pisenti}
\altaffiliation[Present address: ]{IonQ Inc., College Park, MD 20740, USA}
\author{A. Restelli}
\email[]{arestell@umd.edu}
\affiliation{Joint Quantum Institute, University of Maryland and National Institute of Standards and Technology \\ College Park, MD 20742, USA}
\author{J. Scherschligt}
\author{J.A. Fedchak}
\affiliation{Sensor Science Division, National Institute of Standards and Technology \\ Gaithersburg, MD 20899, USA}
\author{G.K. Campbell}
\affiliation{Joint Quantum Institute, University of Maryland and National Institute of Standards and Technology \\ College Park, MD 20742, USA}
\author{S. Eckel}
\affiliation{Sensor Science Division, National Institute of Standards and Technology \\ Gaithersburg, MD 20899, USA}

\date{\today}

\begin{abstract}
We present a design for a radio-frequency driver that leverages telecom amplifiers to achieve high power output and wide bandwidth.
The design consists of two compact printed circuit boards (total area \(< 255~\si{\square\centi\metre}\)), which incorporate power (turn-on) and thermal management to prevent accidental damage to the amplifier circuitry.
Our driver provides \(>1~\si{\watt}\) of output power over a \(10~\si{\mega\hertz}\) to \(1.1~\si{\giga\hertz}\) frequency range, and \(\geq 5~\si{\watt}\) from \(20~\si{\mega\hertz}\) to \(100~\si{\mega\hertz}\).
The driver circuit includes auxiliary components for analog frequency and amplitude modulation (\(\approx 70~\si{\kilo\hertz}\) bandwidth), as well as digital power switching (\(> 30~\si{\decibel}\) of extinction within \(40~\si{\nano\second}\) and final extinction \(> 90~\si{\decibel}\)).
The radio-frequency source can also be digitally switched between an external input and an integrated voltage-controlled oscillator.
Our design is motivated by the need for flexible, inexpensive drivers of optically active devices, such as acousto-optic and electro-optic modulators.
\end{abstract}

\maketitle

\section{Introduction}\label{Sec:Introduction}

Active optical devices are a ubiquitous tool in experimental optics.
In particular, electro-optic (EO) and acousto-optic (AO) devices are used for amplitude, frequency, and phase modulation, as well as beam deflection and polarization control.
AO and EO modulators have found applications in a variety of fields, which include ultrafast laser physics,~\cite{Keller1990} microscopy,~\cite{Morris1994} quantum computing,~\cite{Debnath2016} and spectroscopy.~\cite{Camy1982, Bjorklund1980, Hall1981}
These applications often require analog and digital control of both the amplitude and frequency of the modulator's radio-frequency drive.
For example, analog frequency modulation is necessary for several spectroscopy techniques,~\cite{Negnevitsky2012} fast digital amplitude control is crucial for quantum logic operations,~\cite{Debnath2016} and analog amplitude control is required for stable optical dipole traps.~\cite{Granade2002, Barrett2001}

AO and EO modulators typically operate in the radio-frequency (RF) range and require \(\gtrsim1~\si{\watt}\) of RF power.
Depending on the application, the modulator's RF control circuit may require a frequency tuning range \(\gtrsim100~\si{\mega\hertz}\), power control with \(\gtrsim20~\si{\decibel}\) of dynamic range, and switching times on the order of \(10~\si{\nano\second}\) with \(\gtrsim60~\si{\decibel}\) extinction ratio.
Commercially-available drivers for AO/EO modulators often have a narrow RF bandwidth and typically offer only a subset of the desired RF controls.
Researchers (such as ourselves) often attempt to circumvent the limitations of commercially-available drivers by combining off-the-shelf RF components to produce a custom AO/EO driver.~\cite{ligodriver, frohlich2007}
However, both commercial and custom RF drivers are typically only suitable for specific modulators or applications.

We present a design for a high-flexibility AO/EO device driver that has high output power over a wide frequency bandwidth.
Our circuit uses a custom power sequencing architecture to leverage telecom amplifiers, which allow us to achieve RF signal gain \(\gtrsim30~\si{\decibel}\) and output power \(\gtrsim1~\si{\watt}\) from \(10~\si{\mega\hertz}\) to \(1.1~\si{\giga\hertz}\).
The RF driver consists of two FR4 printed circuit boards (PCBs) with a compact total footprint (\(< 255~\si{\square\centi\metre}\)).
Careful PCB layout keeps the channel temperature of the high-power amplifier well below its maximum rating, even in the absence of forced-air cooling.
Our design incorporates an array of analog and digital controls, which make it compatible with a variety of applications.
The RF driver's output power is continuously tunable with a dynamic range of \(\approx30~\si{\decibel}\) and an analog modulation bandwidth of \(\approx 70~\si{\kilo\hertz}\).
During digital switching, the RF power falls by \(10~\si{\decibel}\) in \(\approx25~\si{\nano\second}\), and by \(>30~\si{\decibel}\) within \(40~\si{\nano\second}\), with a final extinction ratio \(>90~\si{\decibel}\).
The driver also allows digital switching between an integrated voltage-controlled oscillator (VCO), which provides analog frequency tunability, and an external RF source, such as a direct digital synthesizer (DDS) or software-defined radio (SDR).

The RF driver is packaged in a 2U 19" rack enclosure, with the size of the case limited by the driver's linear power supply.\footnote{The PCBs could be adapted to fit in modular electronics crates (\textit{e.g.} Eurocard, NIM). The crate's bus would need to source \(\simeq 1~\si{\ampere}\) at \(24~\si{\volt}\) per installed RF driver.} 
We discuss the circuit design in Section~\ref{Sec:Circuit}.
Design files for both PCBs, including schematics, board layouts, and bills of materials, are available online.~\cite{aogit, gangit}
Section~\ref{Sec:Results} contains our measurements of the RF driver's performance.
We summarize the results and discuss future modifications to our design in Section~\ref{Sec:Conclusion}.
The cost of both fully-populated PCBs is \(\approx\$1000\) for a run of one. 
In a production run of \(40\) RF drivers, the cost per driver, including the rack enclosure and all auxiliary components, was \(\$1085\), which is competitive with commercially-available alternatives. 

\begin{figure*}[t]
\includegraphics[width=\textwidth]{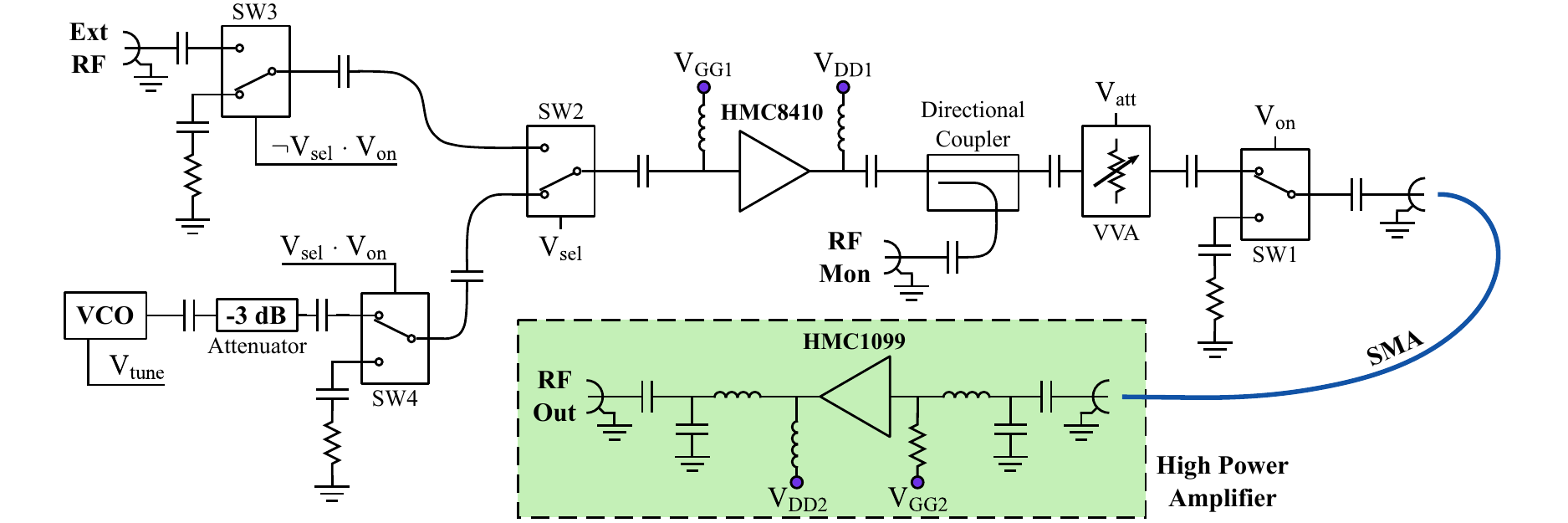}
\caption{\label{Fig:RFCircuit}
Schematic of radio-frequency components.
The RF switch SW2 selects either the VCO or Ext RF connector as the RF signal source.
Switches SW3 and SW4 reduce cross-talk between the two RF signal sources and increase the RF extinction ratio.
The RF signal passes from SW2 into a low-noise, wide-bandwidth pre-amplifier (HMC8410),~\cite{disclaimer} which increases the RF power by \(\approx 20~\si{\decibel}\).
A directional coupler (ADC-20-4) splits off \(\approx 1~\si{\percent}\) of the RF power and sends it to the RF Mon output. 
The RF signal then reaches a voltage-variable attenuator (VVA), which modulates the RF amplitude with a dynamic range of \(\approx 30~\si{\decibel}\).
The switch SW1, in conjunction with SW3/SW4, toggles the RF output power.
A short SMA cable connects the pre-amplified RF signal to a separate high power amplifier PCB (highlighted in green), where the RF signal is amplified to \(\gtrsim 1~\si{\watt}\).
}
\end{figure*}

\section{Circuit Design}\label{Sec:Circuit}

The central features of our design are wide RF amplification bandwidth, high output power, fast digital switching, and high-dynamic-range analog power tuning.
The design requirements were informed by the AO/EO devices commonly employed in our laboratories, which fall into four broad classes: low-power (\(\simeq1~\si{\watt}\)) AO modulators for \(400~\si{\nano\metre}\) to \(800~\si{\nano\metre}\) wavelengths; high-power (\(\simeq5~\si{\watt}\)), low-frequency AO modulators for near-IR wavelengths; resonant free-space EO modulators; and broadband fiber EO modulators.
The combination of the RF bandwidth and low frequency (\(\lesssim100~\si{\mega\hertz}\)) power requirements pose the most significant obstacle to the realization of an RF driver compatible with all these disparate device classes.

\subsection{RF Amplification and Control}\label{Sec:RFdrive}

Figure~\ref{Fig:RFCircuit} is a diagram of our design's RF components.
We arrange the pre-amplifier (HMC8410),~\cite{disclaimer} directional coupler, voltage-variable attenuator (VVA), and rightmost RF switch (SW1) for better dynamic range and extinction.
The component arrangement reduces the attainable RF output power due to the insertion losses in SW1, the VVA, and the directional coupler.
Our component selection represents a compromise between the competing goals of high output power and dynamic range.
The directional coupler picks off \(\approx 1~\si{\percent}\) of the power from the pre-amplifier and sends it to a low-power monitor output.\footnote{Future versions of our design will use two directional couplers placed before SW3 and SW4 (see Figure~\ref{Fig:RFCircuit}) to increase both flexibility and output power.}
The RF components are connected by grounded coplanar waveguides with a designed characteristic impedance of \(50~\si{\ohm}\) at \(1~\si{\giga\hertz}\).
To improve the low-frequency performance, we use \(100~\si{\nano\farad}\) broadband capacitors to AC-couple the RF signal between components.\footnote{Results from a prototype suggest that similar performance can be achieved at lower cost using \(10~\si{\nano\farad}\) standard capacitors with a high self-resonant frequency.}

Two wide-bandwidth, high-gain amplifiers, an HMC1099 and an HMC8410, form the core of our RF driver.~\cite{hmc1099, hmc8410}
The HMC8410 serves as a pre-amplifier, a task for which it is ideally suited due to its low noise figure (\(\lesssim2~\si{\decibel}\)) and gain flatness (\(<1~\si{\decibel}\)) in our frequency range.
The HMC1099 has a saturated output power \(>10~\si{\watt}\) and high power-added efficiency (\(\gtrsim 50~\si{\percent}\) under our operating conditions).
It provides high power amplification and is mounted on a separate PCB (green highlighted region in Figure~\ref{Fig:RFCircuit}) to increase design flexibility.
An inline attenuator can be installed between the PCBs to restrict the maximum output power to prevent damage to delicate optical devices.
Both amplifiers are internally prematched to \(50~\si{\ohm}\) across their full operating bandwidth, which greatly simplifies external impedance matching.
We based the external impedance matching network, including component selection and placement, on the typical application circuits for each amplifier.~\cite{hmc1099, hmc8410, bom}
Unfortunately, both amplifiers lack internal biasing circuitry and also have strict power sequencing requirements.
The gate voltage (\(\text{V}_{\text{GG}}\)), drain voltage (\(\text{V}_{\text{DD}}\)), and RF input must be applied sequentially to each amplifier to prevent damage.
We leave discussion of the details of our power sequencing electronics to Section~\ref{Sec:BiasControl}.

Four RF switches (MASWSS0178) provide digital control of the power output and signal source of the RF driver.~\cite{maswss0178}
Each switch has \(>50~\si{\decibel}\) of isolation and a switching time of \(\approx 20~\si{\nano\second}\).
Two \(5~\si{\volt}\) TTL logic signals, \(\text{V}_{\text{on}}\) and \(\text{V}_{\text{sel}}\), govern the state of the RF switches.
Switch SW1 turns the RF output on or off depending on whether \(\text{V}_{\text{on}}\) is high or low.
Switch SW2 sets the RF source to the VCO when \(\text{V}_{\text{sel}}\) is high, or to the external RF input when \(\text{V}_{\text{sel}}\) is low (\(\neg{\text{V}}_{\text{sel}}\) is high).
The remaining switches, SW3 and SW4, increase the extinction ratio (when \(\text{V}_{\text{on}}\) is low) or reduce bleed-through of the undesired RF source onto the RF output (depending on the state of \(\text{V}_{\text{sel}}\)).

We use a linear-in-\(\si{\decibel}\), voltage-variable attenuator (F2255) for analog adjustment of the driver's RF output power.~\cite{f2255}
The VVA has low insertion loss (\(<1.7~\si{\decibel}\)) and a specified attenuation range of \(33~\si{\decibel}\).
However, gain compression in the high power amplifier limits the dynamic range of the RF driver to \(\approx 30~\si{\decibel}\).
In contrast to its excellent static properties, the F2255 attenuator has a comparatively slow, and poorly specified, response to changes in its control voltage, \(\text{V}_{\text{att}}\).
In applications with lower dynamic range or output power requirements, the VVA can be replaced with a pin-compatible part from another manufacturer, which offer up to \(\approx 10\) times larger amplitude modulation bandwidths.~\cite{vvabw}

We employ a VCO as the RF driver's on-board RF source.
The VCO is controlled and frequency modulated with a tuning voltage, \(\text{V}_{\text{tune}}\).
The RF bandwidth of a VCO is typically much narrower than the bandwidth of the other RF components in our design.
Multiple VCOs could be used in conjunction with a switching network to increase the bandwidth of the on-board RF source,~\cite{Bhandare2019} but the additional components would increase cost and complexity.
We have instead made the RF driver compatible with all \(5~\si{\volt}\) and \(12~\si{\volt}\) VCOs in the CVCO55 series (from Crystek), or the JTOS and ROS series (from Minicircuits).
Most VCOs have sufficient output power to saturate the HMC8410 pre-amplifier.
To reduce saturation of the pre-amplifier, and the associated harmonic distortion, we insert a \(3~\si{\decibel}\) fixed attenuator between the VCO and SW4.
High-frequency VCOs often have a lower output power, so the fixed attenuator can be replaced with a short when driving high-power, high-frequency devices.

\subsection{Analog and Digital Control}\label{Sec:Control}

The RF driver allows both internal and external control of the analog and TTL signals that control its RF output.
Front-panel switches permit the user to chose internal, external, or, for analog signals, summed (internal \(+\) external) control of \(\text{V}_{\text{on}}\), \(\text{V}_{\text{sel}}\), \(\text{V}_{\text{att}}\), and \(\text{V}_{\text{tune}}\).
High-speed optocouplers (TLP2767) detect external TTL signals and provide ground isolation.~\cite{tlp2767}
We configure the optocouplers to be compatible with both \(3.3~\si{\volt}\) and \(5~\si{\volt}\) TTL logic.
A TLP2767 can handle switching waveforms with frequencies as large as \(50~\si{\mega\hertz}\) at \(50~\si{\percent}\) duty cycle.
Analog inputs for amplitude and frequency modulation are sensed pseudo-differentially by instrumentation amplifiers with \(10~\si{\mega\hertz}\) bandwidth (AD8421).
The analog inputs have a \(10~\si{\kilo\ohm}\) input impedance, and a \(50~\si{\ohm}\) resistor isolates the input reference level from circuit ground.
An on-board microcontroller supplies internal setpoints for the analog voltages \(\text{V}_{\text{att}}\) and \(\text{V}_{\text{tune}}\) using a dual-channel digital-to-analog converter (DAC).
Both internal setpoints can be adjusted using a front-panel rotary encoder and LCD display.

Wide-bandwidth op-amps (LM7171) buffer both the internal and external analog control voltages.
These op-amps multiply or divide the analog signals such that the output voltage range of both a typical external voltage source and the DAC will span the input range of the VVA or VCO\@.
In the summed control mode, \(\text{V}_{\text{att}} = \text{V}^{\text{DAC}}_{\text{att}} + \text{V}^{\text{ext}}_{\text{att}}/3\) and \(\text{V}_{\text{tune}} = 3\,(\text{V}^{\text{DAC}}_{\text{tune}} + \text{V}^{\text{ext}}_{\text{tune}})\).
Currently, \(\text{V}_{\text{att}}\) and \(\text{V}_{\text{tune}}\) are only limited by the \(\pm15~\si{\volt}\) power rails of the LM7171 op-amps.
It is therefore possible to exceed the maximum rating of the VVA control voltage and the tuning voltage of certain VCOs when the driver is operated in either the summed or external control modes.
Future versions of the RF driver may integrate Zener diodes on the op-amp outputs to protect the RF components.

In addition to controlling the DAC, the microcontroller is configured to accept triggers on an auxiliary optocoupled TTL line and to control an AD9910 DDS\@.
A DDS shield PCB (based on the AD9910 evaluation board~\cite{ad9910, ddspcb}) can be mounted on to the RF driver, which incorporates extra voltage regulators to power the DDS\@.
The current version of the microcontroller software permits only manual adjustment of the DDS frequency.
The auxiliary TTL trigger enables tabled control of \(\text{V}^{\text{DAC}}_{\text{att}}\), \(\text{V}^{\text{DAC}}_{\text{tune}}\), and all DDS parameters to be implemented in future microcontroller software versions.
The code base and DDS shield could also be adapted to other AD99XX series synthesizers, some of which can generate sinusoidal waveforms over the full RF amplification bandwidth of our RF driver.~\cite{ad9914}

\subsection{Amplifier Power Sequencing and Bias Control}\label{Sec:BiasControl}

The performance of monolithic microwave integrated circuit (MMIC) amplifiers, such as the HMC1099 and HMC8410, depends sensitively on the amplifier's drain current, \(\text{I}_{\text{DD}}\).
The drain current depends on the gate and drain voltages (\(\text{V}_{\text{GG}}\) and \(\text{V}_{\text{DD}}\)) as well as the input RF power.
Excessive drain current can damage the amplifier, so \(\text{V}_{\text{GG}}\), \(\text{V}_{\text{DD}}\), and the RF signal must be applied to the amplifier sequentially.

There are two methods for biasing \(\text{I}_{\text{DD}}\): constant gate voltage and constant drain current.~\cite{Kaya2016}
The constant drain current approach has greater long-term repeatability, but tends to restrict maximum power output because the RF input cannot dynamically pull additional current from the \(\text{V}_{\text{DD}}\) supply.
In the constant gate voltage technique, \(\text{V}_{\text{GG}}\) is set to produce a particular quiescent drain current, \(\text{I}_{\text{DQ}}\), prior to activation of the RF input.
A fixed \(\text{V}_{\text{GG}}\) allows the amplifier to pull additional drain current as the RF input power increases, which reduces gain saturation.
However, the gate voltage that produces the desired \(\text{I}_{\text{DQ}}\) may change as an amplifier ages.
Biasing an amplifier with a constant drain current is generally preferable when an amplifier's optimal \(\text{I}_{\text{DQ}}\) is large enough that
\begin{equation}
\label{Eq:BiasCond}
\text{I}_{\text{DQ}}\text{V}_{\text{DD}} > \text{P}_{\text{sat}},
\end{equation}
where \(\text{P}_{\text{sat}}\) is the saturated output power.
Condition~(\ref{Eq:BiasCond}) is valid for the HMC8410 pre-amplifier, but not the HMC1099 high-power amplifier.
Consequently, the RF driver employs both drain current biasing procedures.

We control the HMC8410 pre-amplifier using an active bias controller (HMC920LP5E), which also ensures proper sequencing of \(\text{V}_{\text{GG1}}\) and \(\text{V}_{\text{DD1}}\) during start up and shut down (see Figure~\ref{Fig:RFCircuit}).
Our bias control circuit is an amalgamation of example circuits from the HMC920LP5E and HMC8410 datasheets.~\cite{hmc8410, hmc920}
To increase the pre-amplifier's saturated output power, we bias the HMC8410 with \(\text{V}_{\text{DD1}} = 6.4~\si{\volt}\) and \(\text{I}_{\text{DD1}} = 75~\si{\milli\ampere}\).
Operating the HMC8410 under these bias conditions, at the high end of the permitted range, slightly degrades its noise figure.
In applications with lower power requirements, the noise figure can be improved by reconfiguring the active bias control circuit.

\begin{figure}[t]
\includegraphics[width=\linewidth]{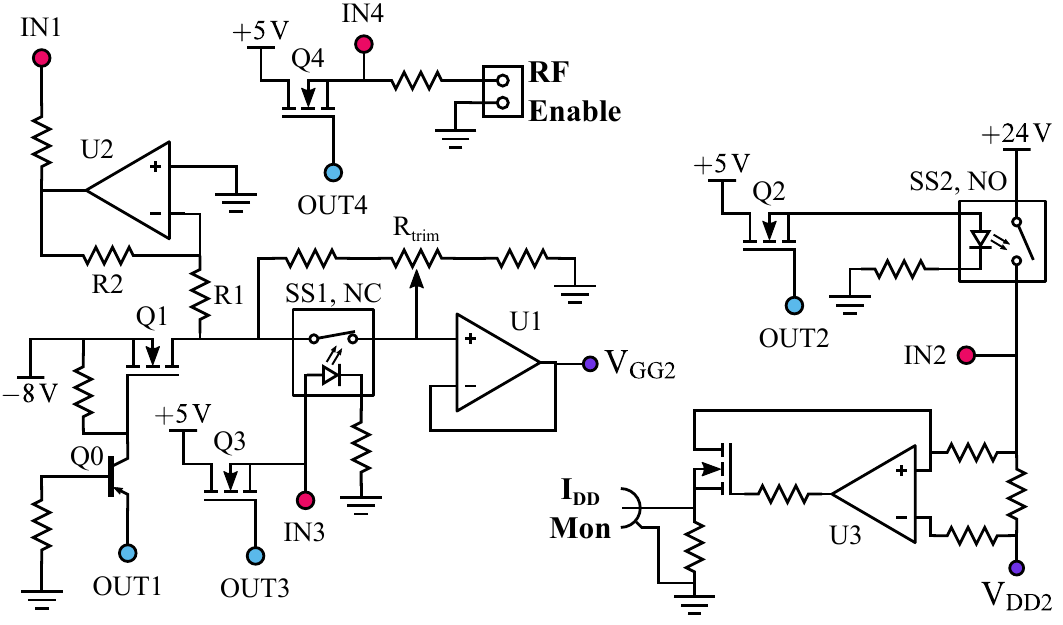}
\caption{\label{Fig:Bias}
Simplified schematic of the \(\text{I}_{\text{DD}}\) bias controller for the high-power amplifier.
Blue-filled (red-filled) circles denote the output (input) pins of a LTC2924 power sequencer.
Purple-filled circles mark the gate and drain voltage of the HMC1099 amplifier (see Figure~\ref{Fig:RFCircuit}).
The power sequencer consecutively activates OUT1 through OUT4 to put transistors Q1 through Q4 into conduction.
As each transistor turns on, the LTC2924 waits for the associated input (IN1 to IN4) to reach a resistively programmed threshold value before activating the next output (the threshold setting voltage dividers are omitted for clarity).
The transistors Q1 to Q3 successively set \(\text{V}_{\text{GG2}}\) to \(-8~\si{\volt}\); raise \(\text{V}_{\text{DD2}}\) to \(24~\si{\volt}\); and switch \(\text{V}_{\text{GG2}}\) to a value determined by \(\text{R}_{\text{trim}}\).
The \(\text{I}_{\text{DD}}\) Mon output allows us to adjust \(\text{R}_{\text{trim}}\) to obtain a quiescent drain current \(\text{I}_{\text{DQ2}}=100~\si{\milli\ampere}\).
Transistor Q4 actuates an optocoupled solid-state relay on the pre-amplifier PCB via the RF Enable output.
The solid-state relay locks \(\text{V}_{\text{on}}\) low until both amplifiers' gate and drain voltages have stabilized.
}
\end{figure}

A power supply sequencer (LTC2924) governs the gate and drain voltages for the high-power amplifier.~\cite{ltc2924}
The LTC2924 can drive the gates of external N-channel MOSFETs to sequence up to four external supply voltages.
Due to its limited output voltage, the LTC2924 can normally only control positive power supplies with output \(\leq 5~\si{\volt}\).
The HMC1099 requires \(\text{V}_{\text{DD2}}\geq 24~\si{\volt}\) during normal operation and \(\text{V}_{\text{GG2}}\) as low as \(-8~\si{\volt}\) during power sequencing.

Figure~\ref{Fig:Bias} shows the central external circuitry that we use to adapt the LTC2924 for control of both negative-voltage and large positive-voltage power supplies.
Blue (red) circles denote the power sequencer's output (input) pins, OUT1 through OUT4 (IN1 through IN4).
When we enable the LTC2924, it sets OUT1 high (\(\approx 10~\si{\volt}\)), which puts the PNP transistor Q0 into conduction and so pulls the gate of MOSFET Q1 to ground.
As Q1 goes into conduction, the buffering op-amp U1 pulls \(\text{V}_{\text{GG2}}\) down to \(-8~\si{\volt}\) to pinch off the drain current.
The inverting op-amp U2 converts \(\text{V}_{\text{GG2}}\) into a positive voltage, which the power sequencer can sense at IN1.
The LTC2924 waits until \(\text{V}_{\text{GG2}}\) reaches a threshold value, \(\text{V}_{\text{th}} = -0.61~\si{\volt}\times(R1/R2)\), before activating OUT2.
OUT2 actuates a \(5~\si{\volt}\) signal that closes the normally-open (NO), optocoupled solid-state relay SS2 and brings \(\text{V}_{\text{DD2}}\) up to \(24~\si{\volt}\).
Pin IN2 senses the HMC1099 drain voltage through a resistive divider (not shown in Figure~\ref{Fig:Bias}).
Once \(\text{V}_{\text{DD2}}\) stabilizes, the power sequencer raises OUT3 to open the normally-closed (NC) relay SS1.
The gate voltage then switches from \(-8~\si{\volt}\) to a value set by the high-precision potentiometer \(\text{R}_{\text{trim}}\).
The op-amp U3 monitors the drain current, \(\text{I}_{\text{DD2}}\), and allows us to adjust \(\text{R}_{\text{trim}}\) to achieve the recommended quiescent drain current, \(\text{I}_{\text{DQ2}}=100~\si{\milli\ampere}\).

A front-panel rocker switch enables both the pre-amplifier's active bias controller and high-power amplifier's power sequencer.
To prevent RF-induced damage to both amplifiers during power sequencing, two optocoupled solid-state relays lock \(\text{V}_{\text{on}}\) low when the rocker switch is off.
Once both amplifiers are on, the status output of the active bias controller and OUT4 of the power sequencer switch the two relays, which permits the RF signal to be enabled.
Both controllers reverse their power sequences when the rocker switch is flipped off, so both amplifiers are fully protected from damage.

During normal operation, the rocker switch should be in the off position whenever the driver's AC mains are switched on or off.
We also tested the operation of the power sequencers when the AC mains are improperly toggled while the rocker switch was turned on.
To increase the potential for damage to the amplifiers, we applied \(1~\si{\milli\watt}\) to the external RF input and set \(\text{V}_{\text{on}}\) high during the test.
After \(100\) improper power sequencing cycles there was no degradation of the RF driver's performance.
Note that the amplifiers are not auto-terminated, so they can still be damaged if the power sequencers are activated when no output load is connected.

\subsection{Thermal Management}\label{Sec:Thermal}

Even though the high-power amplifier has high power-added efficiency, it still generates \(\simeq 5~\si{\watt}\) of heat at maximum RF output power.
We use a \(40.6~\si{\milli\metre}\times40.6~\si{\milli\metre}\times13.3~\si{\milli\metre}\) ball grid array (BGA) heatsink and a \(24~\si{\volt}\) DC fan to dissipate the heat.
The BGA heatsink is installed on the backside of the amplifier PCB using high-thermal-conductivity epoxy (\(K\approx1.0~\si{\watt\per\metre\per\kelvin}\)).
The area beneath the heatsink has no solder mask and the only traces in this region, on any layer of the PCB, are power or signal connections to the HMC1099.
More than 100 vias provide heat conduction from the bottom layer of the PCB, where the amplifier is attached, to the top layer, with the BGA heatsink.
The amplifier PCB is mounted horizontally in the RF driver box with the fan attached the side of the box and \(\approx5~\si{\centi\metre}\) from the heatsink.
When we operate the RF driver with forced-air cooling, the HMC1099 remains within \(20~\si{\degreeCelsius}\) of ambient regardless of the RF input power (see Section~\ref{Sec:Results}, Figure~\ref{Fig:Harmonics}).

The acoustic noise of the DC fan may be unacceptable in certain laboratory environments.
With the fan disabled, the BGA heatsink equilibrates at \(\approx 30~\si{\degreeCelsius}\) above ambient with no RF input and \(\text{I}_{\text{DQ2}}=100~\si{\milli\ampere}\).
When we apply \(10~\si{\milli\watt}\) to the external RF input, the BGA heatsink stabilizes at \(\approx 45~\si{\degreeCelsius}\) above ambient, which corresponds to a heat load of \(\approx 5~\si{\watt}\).
If we assume that the copper acts only to spread heat within a PCB layer (\textit{i.e.} we neglect the presence of the vias) and estimate that the thermal conductivity of FR4 is \(0.25~\si{\watt\per\metre\per\kelvin}\), then the thermal resistance from the heatsink to the amplifier's thermal pad is \(\approx 4.3~\si{\degreeCelsius\per\watt}\).
The junction-to-pad thermal resistance of the HMC1099 is \(11.2~\si{\degreeCelsius\per\watt}\), so the junction temperature is \(\lesssim 140~\si{\degreeCelsius}\) under passive cooling (the absolute maximum junction temperature for the HMC1099 is \(225~\si{\degreeCelsius}\)).
However, the absolute maximum case temperature for the HMC1099 is only \(85~\si{\degreeCelsius}\).
Our measurements suggest that the amplifier case temperature could approach the maximum rating in environments with ambient temperatures \(>20~\si{\degreeCelsius}\).
Laboratories with poor temperature stability should either use forced-air cooling (as we suggest above) or a custom copper heatsink to improve the thermal performance.

\section{Results}\label{Sec:Results}

\begin{figure}
\includegraphics[width=\linewidth]{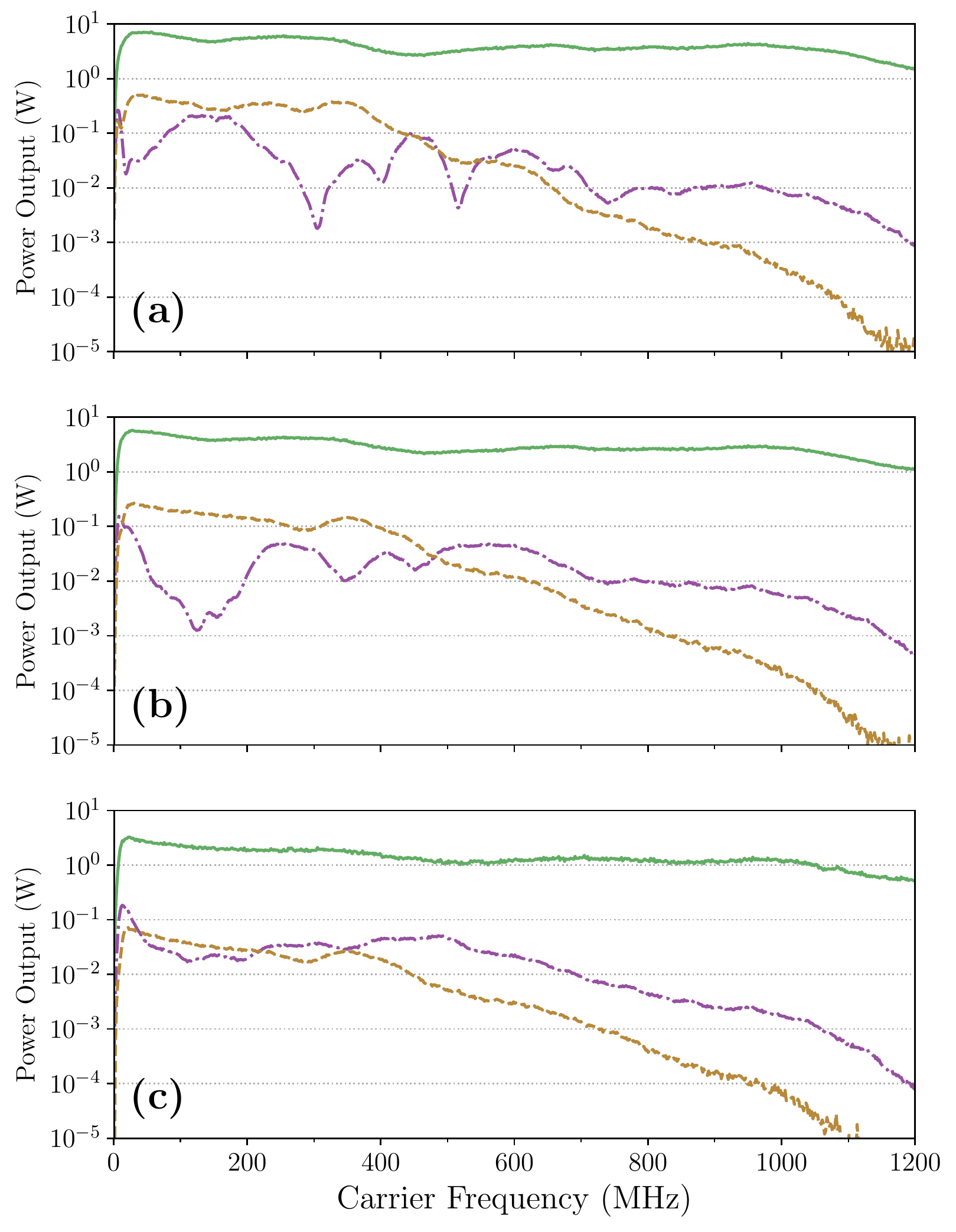}
\caption{\label{Fig:Harmonics}
Power output in the 1st, 2nd and 3rd harmonics (green solid, purple dash-dotted, and gold dashed lines, respectively) as a function of carrier frequency.
The RF power applied to the external input is \(\{10~\si{\milli\watt}, 3.2~\si{\milli\watt}, 1~\si{\milli\watt}\}\) in subplot \{(a), (b), (c)\}. 
}
\end{figure}

We show the RF power output of our design in Figure~\ref{Fig:Harmonics}.
We measured the RF power in the 1st, 2nd, and 3rd harmonic as function of frequency using a Rohde \& Schwarz FSV scalar network analyzer with a controllable SMB100A tracking generator.
The RF output of the tracking generator was connected to the external RF input of a test driver, which was attached to the network analyzer through a \(30~\si{\decibel}\) attenuator and a DC block.
The transmission losses of the external RF components and cables were calibrated out by the network analyzer.
For a \(1~\si{\milli\watt}\) input power, our driver outputs \(>1~\si{\watt}\) for input frequencies from \(10~\si{\mega\hertz}\) to \(1~\si{\giga\hertz}\) (see Figure~\ref{Fig:Harmonics}~(c)).
When we increase the input power to \(3.2~\si{\milli\watt}\) (Figure~\ref{Fig:Harmonics}~(b)) or \(10~\si{\milli\watt}\) (Figure~\ref{Fig:Harmonics}~(a)), the driver can output \(>1~\si{\watt}\) at frequencies up to \(1.2~\si{\giga\hertz}\) and \(\geq 5~\si{\watt}\) from in the \(20~\si{\mega\hertz}\) to \(100~\si{\mega\hertz}\) frequency range.
The increase in carrier output power is accompanied by an increase in harmonic distortion, particularly at frequencies \(\lesssim 400~\si{\mega\hertz}\) and in the 3rd harmonic.
The harmonic distortion (referred to the carrier) is always \(< -10~\si{\decibel}\) inside the RF driver's operating bandwidth and agrees with the 2nd harmonic data in the HMC1099 datasheet.~\cite{hmc1099, otheramp}

Commercially-available RF drivers that specify harmonic distortion typically report harmonic content (referred to the carrier) \(< -15~\si{\decibel}\) or \(< -20~\si{\decibel}\).
Our driver meets the less-stringent specification at low input power (Figure~\ref{Fig:Harmonics}~(c)) except for frequencies \(\lesssim 30~\si{\mega\hertz}\) and in the range from \(400~\si{\mega\hertz}\) to \(525~\si{\mega\hertz}\).
The driver also meets the more-stringent specification at all input powers for frequencies \(> 700~\si{\mega\hertz}\).
When driving narrow-bandwidth AO or EO devices, the harmonic distortion is unlikely to be problematic because higher harmonics will be filtered by the device itself.
External filtering may be beneficial for applications involving wide-bandwidth devices, such as fiber EO modulators, especially when the bandwidth of the device exceeds the bandwidth of the RF driver.
However, broadband AO/EO devices typically require less RF drive power and have non-linear response to weak RF drives.
These device characteristics tend to reduce both the amount of harmonic distortion and its effect.
We estimate that, for a broadband fiber phase modulator driven at a typical half-wave voltage (\(\simeq 5~\si{\volt}\)) and a harmonic distortion (referred to the carrier) of \(-13~\si{\decibel}\), the fractional power in the first-order sidebands associated with the dominant harmonic will be \(< 10~\si{\percent}\).
For lower modulations depths (\textit{i.e.} when only the first-order sidebands contain appreciable power), the fractional power in the first-order sidebands associated with the dominant harmonic will be \(\lesssim 1~\si{\percent}\).

\begin{figure}[t]
\includegraphics[width=\linewidth]{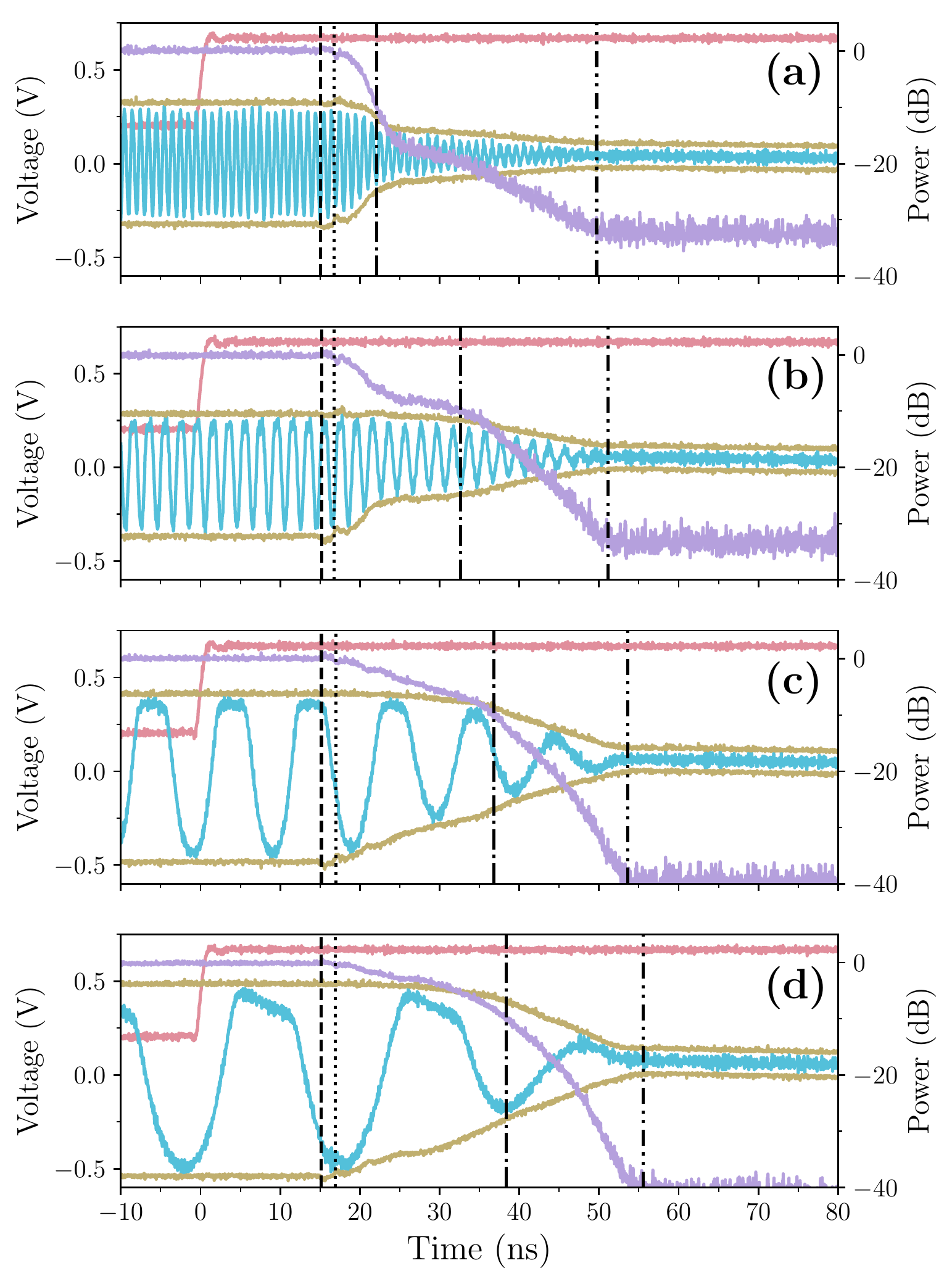}
\caption{\label{Fig:Switching}
Digital switching performance of the RF driver at input frequencies of \(1~\si{\giga\hertz}\) (a), \(500~\si{\mega\hertz}\) (b), \(100~\si{\mega\hertz}\) (c), and \(50~\si{\mega\hertz}\) (d).
Each subplot shows the RF waveform (blue), RF envelope (gold), and TTL signal (red) on the left axis.
The relative peak-to-peak power, \(P_{\text{pp}}\), (see Equation~\ref{Ppeak}) is plotted against the right axis in purple.
The TTL signal is vertically offset for clarity.
The vertical dashed line marks the initial response of the RF output and the vertical \{dotted, dash-dotted, dash-dot-dotted\} lines show the time at which the RF output falls to \(\{90~\si{\percent}\), \(10~\si{\percent}\), \(<0.1~\si{\percent}\}\) of its initial value. 
For all these data, the VVA was set for minimum attenuation.
The harmonic distortion that is apparent in each subplot is consistent with the measurements in Figure~\ref{Fig:Harmonics}.
}
\end{figure}

We evaluated the switching time of the RF driver using a DSA72004B \(20~\si{\giga\hertz}\) bandwidth oscilloscope.
A DG4162 waveform generator drove the digital power control input of the RF driver with a \(5~\si{\volt}\) TTL signal at a frequency of \(100~\si{\kilo\hertz}\).
Figure~\ref{Fig:Switching} displays the TTL signal, RF waveforms, and RF envelopes acquired at several input frequencies.
Using the upper and lower envelopes (\(\text{V}_{\text{upper}}(t)\) and \(\text{V}_{\text{lower}}(t)\)), we compute the relative peak-to-peak power
\begin{equation}
\label{Ppeak}
P_{\text{pp}}(t)=\frac{{(\text{V}_{\text{upper}}{(t)}-\text{V}_{\text{lower}}{(t)})}^{2}}{{(\text{V}_{\text{upper}}{(0)}-\text{V}_{\text{lower}}{(0)})}^{2}},
\end{equation}
which we plot on the righthand axis in Figure~\ref{Fig:Switching}.
The peak-to-peak power begins to decrease \(\approx 15~\si{\nano\second}\) after the arrival of the TTL signal edge (vertical dashed line in Figure~\ref{Fig:Switching}).
The propagation delay is consistent with the specifications of the optocouplers and TTL logic chips that drive SW1 through SW3 (the delay due to mismatched cable lengths is \(\leq 1~\si{\nano\second}\)).
Once the RF switches begin to turn off, \(P_{\text{pp}}\) diminishes by \(10~\si{\decibel}\) within \(25~\si{\nano\second}\) (vertical dash-dotted line in Figure~\ref{Fig:Switching}) and it falls below the noise floor of our measurement technique, \(\leq -30~\si{\decibel}\), within \(40~\si{\nano\second}\) (vertical dash-dot-dotted line in Figure~\ref{Fig:Switching}).
The switching time is consistent with the specification of the RF switches and is comparable to rise/fall times of commercially-available drivers (the fall time of our driver is the distance between the vertical dotted and dash-dotted lines in Figure~\ref{Fig:Switching}).
Changing from a \(5~\si{\volt}\) to a \(3.3~\si{\volt}\) TTL input signal increases the propagation delay by several nanoseconds, but does not increase the switch-off time.
Note that the root-mean-square power, which is the true figure of merit, falls at least as quickly as \(P_{\text{pp}}\), so the switching times that we quote represent an upper bound on the performance of the RF driver in laboratory applications.
The full turn-on time for the RF output is approximately a factor of two longer than the turn-off time (the positive voltage envelope responds less quickly during turn-on), which limits the maximum digital modulation frequency to \(\lesssim10~\si{\mega\hertz}\) (at \(50~\si{\percent}\) duty cycle).
The RF switches have a symmetric switching time specification, so the asymmetry of the rise and fall times of the RF driver is likely caused by asymmetric response of the high-power amplifier.

\begin{figure}
\includegraphics[width=\linewidth]{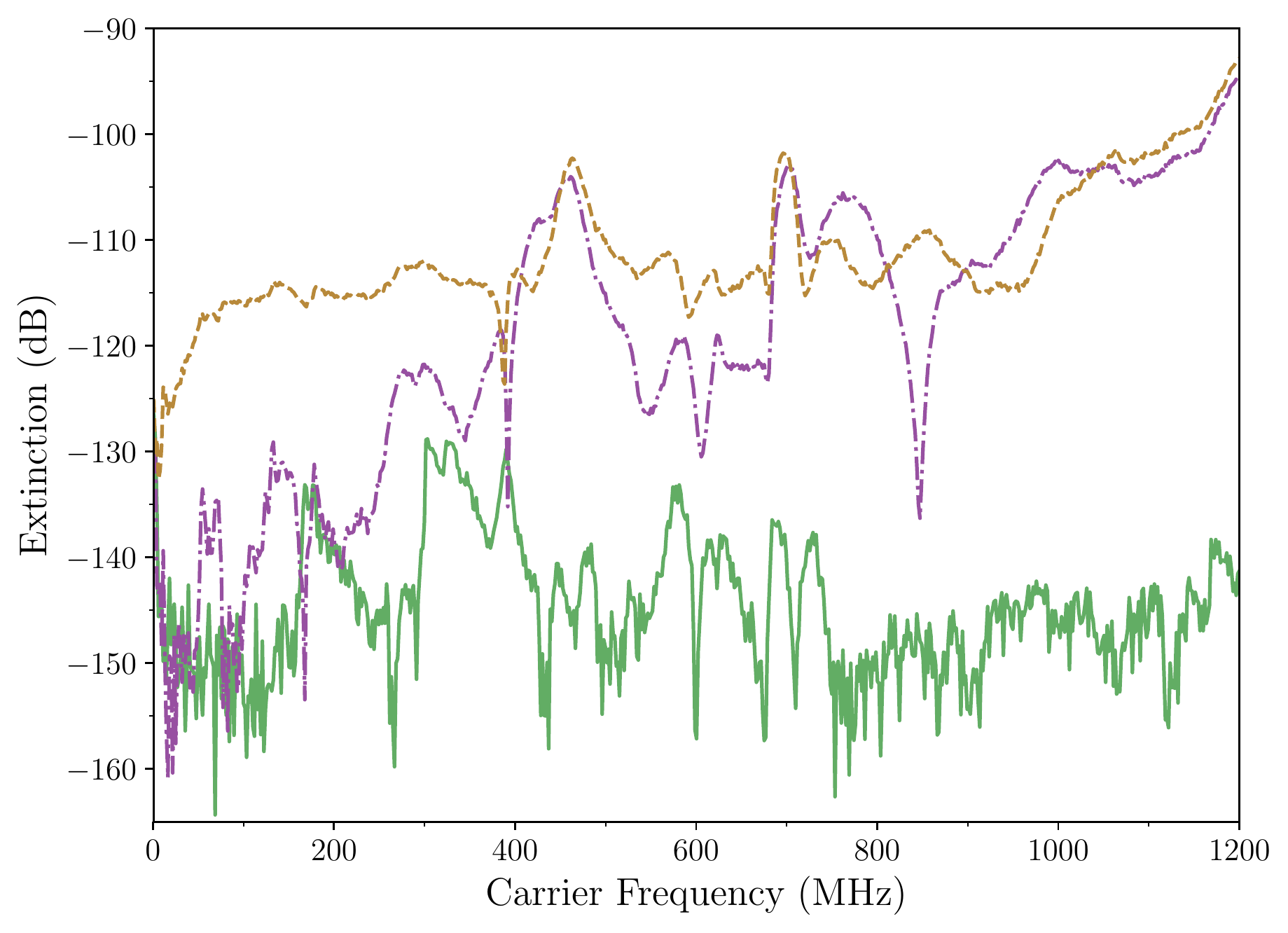}
\caption{\label{Fig:Extinction}
Extinction of the RF output as a function of frequency at \(1~\si{\milli\watt}\) input power.
Each trace is the difference between the carrier output power with \(\text{V}_{\text{on}}\) low and \(\text{V}_{\text{on}}\) high (the solid green curve in Figure~\ref{Fig:Harmonics} (c)).
The gold dashed (purple dash-dotted) curves show the extinction with the minimum (maximum) VVA attenuation setting.
The green solid line represents the noise floor of the measurement.
We acquired the \(\text{V}_{\text{on}}\) low portion of the data for the noise floor measurement with the RF input to the driver terminated, but with the tracking generator turned on and the VVA attenuation maximized.
}
\end{figure}

\begin{figure}
\includegraphics[width=\linewidth]{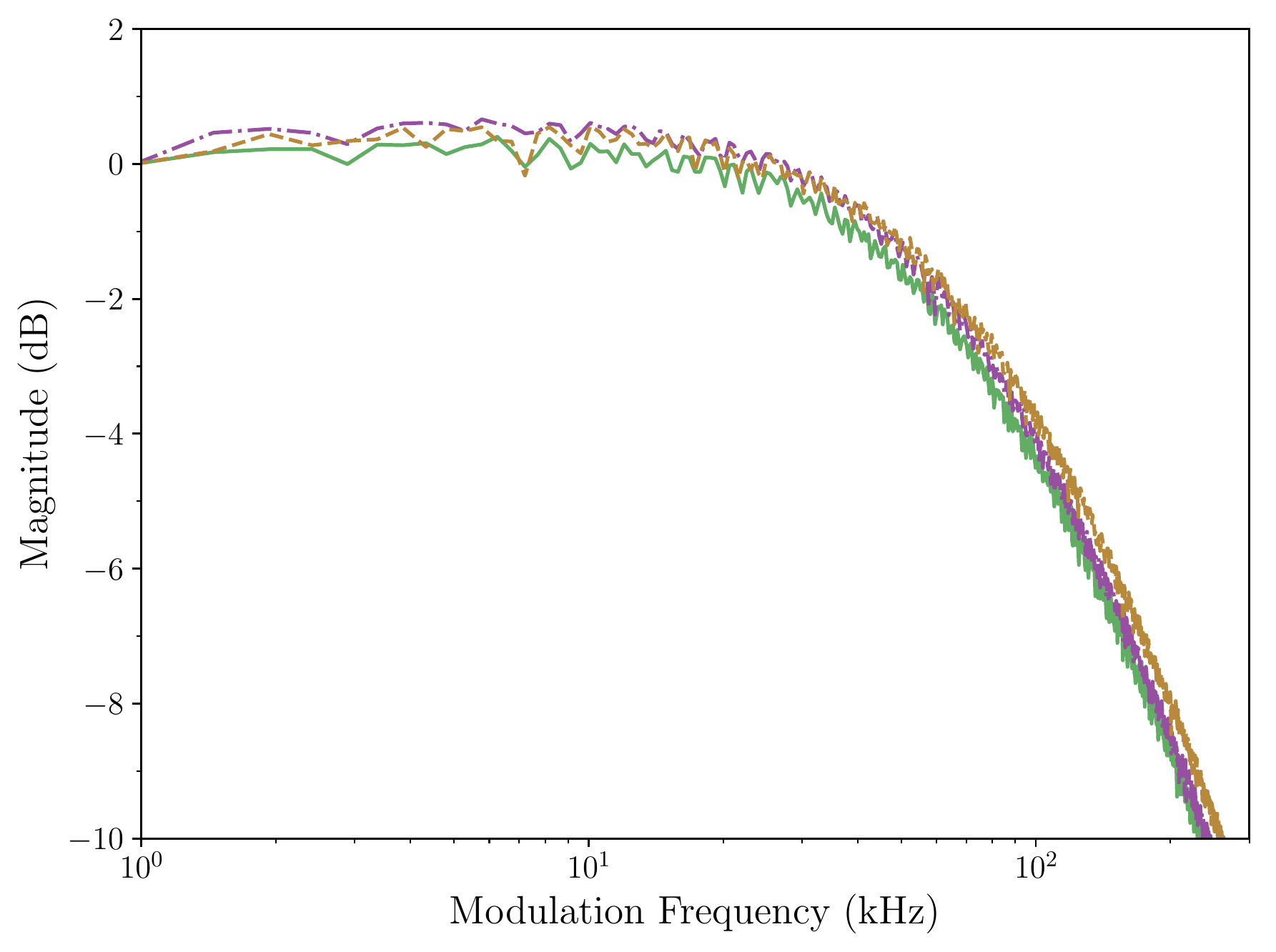}
\caption{\label{Fig:AM}
Relative RF output amplitude modulation as a function of modulation frequency.
The green solid, gold dashed, and purple dash-dotted lines show the relative magnitude of the RF output response for sinusoidal peak-to-peak modulation depths of \(100~\si{\milli\volt}\), \(200~\si{\milli\volt}\), and \(400~\si{\milli\volt}\), respectively.
The \(\{100~\si{\milli\volt}\), \(200~\si{\milli\volt}\), \(400~\si{\milli\volt}\}\) peak-to-peak modulation depth corresponds to a peak-to-peak RF power variation of \(\{3~\si{\decibel}\), \(6~\si{\decibel}\), \(12~\si{\decibel}\}\) at DC\@. 
}
\end{figure}

We investigated the final RF extinction ratio using the FSV network analyzer.
The tracking generator sent \(1~\si{\milli\watt}\) of RF power to the driver and the network analyzer recorded the RF output with \(\text{V}_{\text{on}}\) held low.
We measured the RF power output with the VVA set for both minimal and maximal attenuation.
The extinction is the difference between the measured RF output with \(\text{V}_{\text{on}}\) low and \(\text{V}_{\text{on}}\) high.
To calculate an extinction noise floor, we also acquired the RF output power with the driver's RF input terminated, \(\text{V}_{\text{on}}\) held low, the VVA set for maximum attenuation, and the tracking generator turned on.\footnote{We find that RF pickup from the tracking generator, even when it is not attached to the RF driver, is an important contribution to the measured RF spectrum.}
Figure~\ref{Fig:Extinction} shows the results of our extinction measurements.
Regardless of the VVA attenuation setting, we observe extinction ratios \(>90~\si{\decibel}\) over the full RF bandwidth of the RF driver, and \(>110~\si{\decibel}\) for frequencies \(<400~\si{\mega\hertz}\).
The noise floor for our measurement varies with frequency, but always corresponds to an extinction ratio \(>125~\si{\decibel}\).
The extinction ratios that we observe, to the best of our knowledge, surpass the specification of any commercially-available driver by at least \(20~\si{\decibel}\).

The modulation bandwidth of the VVA is not well specified and RF amplitude modulation bandwidth is important in several applications (\textit{e.g.} laser intensity stabilization~\cite{Granade2002}).
To study the AM bandwidth of the RF driver, we set \(\text{V}^{\text{DAC}}_{\text{att}}\) to the center of the VVA's linear-in-\(\si{\decibel}\) range and drove \(\text{V}^{\text{ext}}_{\text{att}}\) with a sinusoid.
The network analyzer recorded the amplitude of the modulation of the RF output as a function of the modulation frequency.
Because the FSV is a scalar network analyzer, we cannot measure the full amplitude modulation transfer function of the RF driver.
Figure~\ref{Fig:AM} shows the relative magnitude of the amplitude modulation response for several modulation depths.
The amplitude modulation bandwidth increases slightly with increasing modulation depth and the \(3~\si{\decibel}\) point for the lowest modulation depth is \(\approx 70~\si{\kilo\hertz}\).\footnote{When using the RF driver to intensity stabilize laser beams, we have observed intensity locking bandwidths \(\gtrsim 60~\si{\kilo\hertz}\). The locking performance suggests that the VVA has reasonable phase margin at the \(3~\si{\decibel}\) point.}
Our measurement is in reasonable agreement with the small-signal modulation bandwidth suggested by the F2255 VVA datasheet (\(\approx 65~\si{\kilo\hertz}\)).
In applications where wider modulation bandwidths are necessary, the VVA can be replaced with a pin-compatible part with \(\approx 10\) times faster response.~\cite{vvabw}

\begin{figure}[t]
\includegraphics[width=\linewidth]{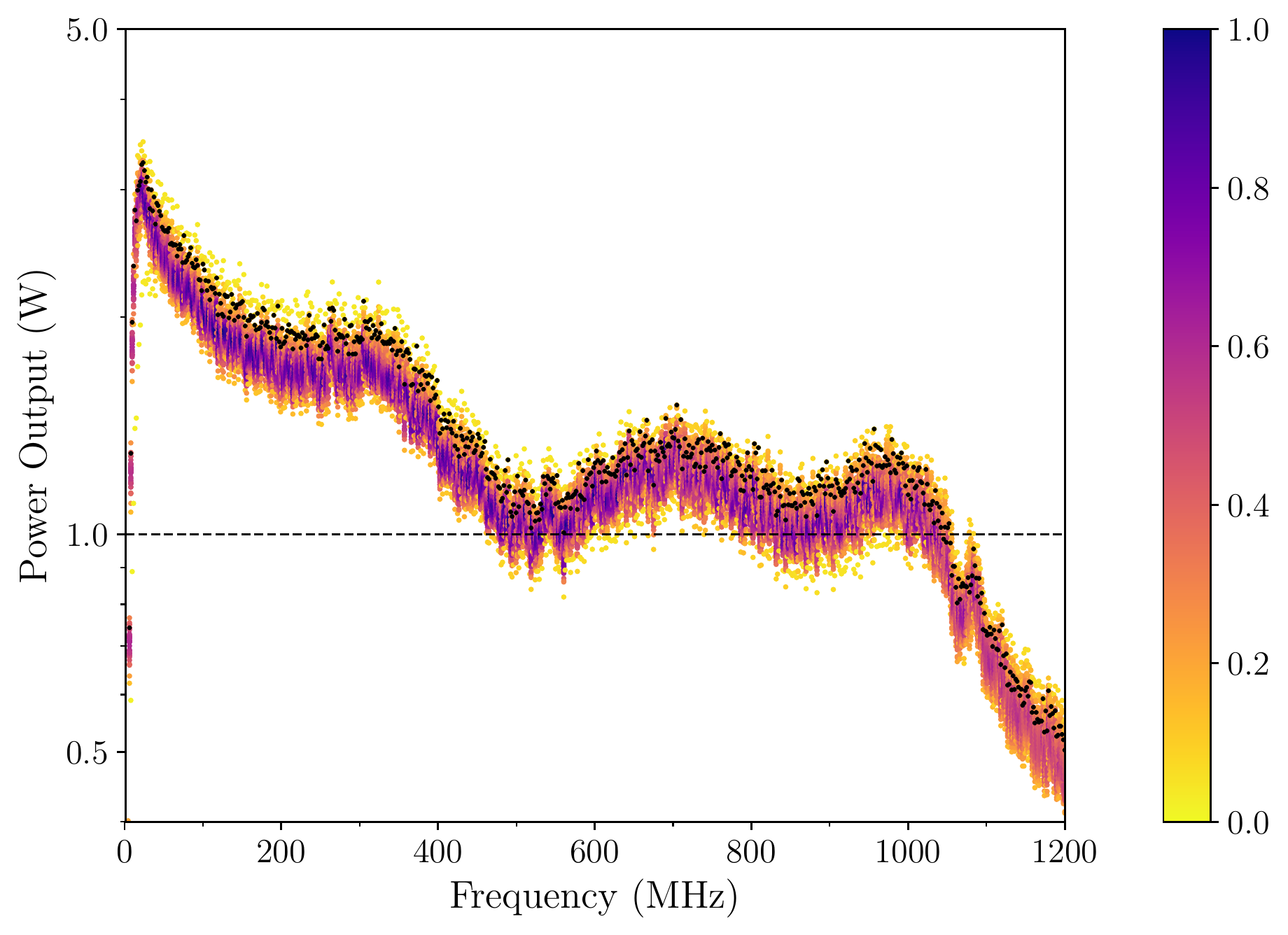}
\caption{\label{Fig:Dist}
Output power for RF drivers from the initial production run.
The external RF input power for these data is \(1~\si{\milli\watt}\).
The output powers are binned by frequency and colored according to the normally-distributed probability of occurrence for a given output power.
The color scale is normalized to the most probable output power in each bin.
Black dots show the output power for the test driver that we used to acquire the data in Section~\ref{Sec:Results}.
The black dashed horizontal line is a guide to the eye.
The device-to-device variation in the output power is within the specification of the amplifiers and is exaggerated by our measurement's resolution bandwidth, which is \(10~\si{\kilo\hertz}\).
}
\end{figure}

\section{Conclusion}\label{Sec:Conclusion}

\begin{table}[b]
\caption{\label{Tab:VCO}
Maximum power output of the production run RF drivers when using the integrated VCO as the RF source.
Voltage-controlled oscillators with nominal frequencies of \(\{40~\si{\mega\hertz}\), \(80~\si{\mega\hertz}\), \(200~\si{\mega\hertz}\), \(400~\si{\mega\hertz}\), \(800~\si{\mega\hertz}\}\) were installed in \(\{4\), \(22\), \(6\), \(1\), \(3\}\) drivers. 
We report the mean and standard deviation of the maximum output power for each VCO frequency class with the \(3~\si{\decibel}\) fixed attenuator installed before the pre-amplifier (see Figure~\ref{Fig:RFCircuit}).
The larger relative output power variation at \(800~\si{\mega\hertz}\) is caused by output power differences between VCOs (we used two \(800~\si{\mega\hertz}\) VCO models, due to product discontinuation).
}
\begin{ruledtabular}
\begin{tabular}{cc}
VCO Frequency & Output Power \\
\hline 
\(40~\si{\mega\hertz}\) & \(5.3(1)~\si{\watt}\)\T \\
\(80~\si{\mega\hertz}\) & \(4.6(2)~\si{\watt}\) \\
\(200~\si{\mega\hertz}\) & \(3.7(1)~\si{\watt}\) \\
\(400~\si{\mega\hertz}\) & \(1.9~\si{\watt}\) \\
\(800~\si{\mega\hertz}\) & \(1.4(2)~\si{\watt}\) \\
\end{tabular}
\end{ruledtabular}
\end{table}

We have designed and characterized an RF driver for AO and EO devices.
The driver achieves high output power and wide RF bandwidth using telecom amplifiers, which are fully protected during power cycles by a combination of custom and commercial power sequencing electronics.
The driver includes flexible controls for AM, FM, and digital switching of the RF output.
In addition to the test data on a single driver that we relate in Section~\ref{Sec:Results}, we also have more limited data from a production run of \(40\) drivers.
We present the power output performance of drivers from the production run in Figure~\ref{Fig:Dist} and Table~\ref{Tab:VCO}.
Our design could be modified to increase digital switching speed, analog AM bandwidth, or extinction ratio at the expense of power output.
Our power sequencing and bias control electronics can be adapted for other amplifiers and applications.
All design materials are available online for others to use.~\cite{aogit, gangit}

\begin{acknowledgments}
The authors thank J. Gardner and K. Douglass for their careful reading of the manuscript.
Our work was partially supported by the Office of Naval Research, and the National Science Foundation through the Physics Frontier Center at the Joint Quantum Institute.
DSB acknowledges support from the National Research Council Postdoctoral Research Associateship Program.
\end{acknowledgments}

\bibliography{AODriver}

\begin{thebibliography}{34}
\bibitem [{dis()}]{disclaimer}%
  \BibitemOpen
  \href@noop {} {}\bibinfo {note} {The identification of commercial products is
  for information only and does not imply recommendation or endorsement by the
  National Institute of Standards and Technology}\BibitemShut {NoStop}%
\bibitem [{oth()}]{otheramp}%
  \BibitemOpen
  \href@noop {} {}\bibinfo {note} {Our power sequencing electronics could be
  adapted for other high-power amplifiers with superior 2nd harmonic
  distortion. These amplifiers require $\text{V}_{\text{DD}} > 24~\si{\volt}$,
  so an additional power supply would be necessary.}\BibitemShut {Stop}%
\end{thebibliography}

\end{document}